\documentclass[preprint,10pt]{article}
\usepackage{macro}

\usepackage{amsmath}
\usepackage{amssymb}
\usepackage{bm}

\usepackage{graphicx}
\usepackage{afterpage}
\usepackage{subfigure}

\newcommand{\avbr}[1]{\left\langle {#1} \right\rangle}
\newcommand{\cbr}[1]{\left( {#1} \right)}

\newcommand{\sbr}[1]{\left[ {#1} \right]}

\newcommand{\fr}[2]{\frac{#1}{#2}}

\newcommand{\vecs}[2]{\mathbf{#1}_{#2}}

\newcommand{\Refcite}[1]{Ref.~\citen{#1}}

\begin{document}

\title{Collapse dynamics of copolymers in a poor
solvent: Influence of hydrodynamic interactions and chain sequence}

\author{
Tri Thanh Pham}
\affiliation{Department of Chemical Engineering, Monash University,
VIC-3800, Melbourne, Australia \\
Max Planck Institute for Polymer Research,
Ackermannweg 10, D-55128 Mainz, Germany
}
\author{Burkhard D\"{u}nweg} 
\affiliation{Max Planck Institute for Polymer Research,
Ackermannweg 10, D-55128 Mainz, Germany
}
\author{J. Ravi Prakash\footnote{Corresponding author.
Email: ravi.jagadeeshan@monash.edu}
}
\affiliation{Department of Chemical Engineering, Monash University,
VIC-3800, Melbourne, Australia 
}

\begin{abstract}
  We investigate the dynamics of the collapse of a single copolymer
  chain, when the solvent quality is suddenly quenched from good to
  poor. We employ Brownian dynamics simulations of a bead-spring chain
  model and incorporate fluctuating hydrodynamic interactions via the
  Rotne-Prager-Yamakawa tensor. Various copolymer architectures are
  studied within the framework of a two-letter HP model, where
  monomers of type H (hydrophobic) attract each other, while all
  interactions involving P (polar or hydrophilic) monomers are purely
  repulsive. The hydrodynamic interactions are found to assist the
  collapse. Furthermore, the chain sequence has a strong influence on
  the kinetics and on the compactness and energy of the final
  state. The dynamics is typically characterised by initial rapid
  cluster formation, followed by coalescence and final rearrangement
  to form the compact globule.  The coalescence stage takes most of
  the collapse time, and its duration is particularly sensitive to the
  details of the architecture.  Long blocks of type P are identified
  as the main bottlenecks to find the globular state rapidly.
\end{abstract}


\maketitle

\section{\label{sec:Intro}Introduction}

The dynamics of the conformational change of a single polymer chain
from a swollen coil to a collapsed globule due to a sudden change in
the solvent quality has received much attention over the past decades
\cite{Nishio79,deGennesBook79,deGennes85,Buguinetal96}. One of the
primary reasons for the continued interest in polymer collapse
dynamics is because of its qualitative similarity with the folding
transitions seen in proteins \cite{Creigton94}. It is believed that
understanding the collapse transition will enable us to obtain better
insight into the conformational transitions in many complex biological
systems, and to understand how other bio-molecules such as DNA react
to a change in their environment.

Since the folding or collapse process of a chain mainly occurs in an
environment surrounded by solvent molecules, it is now widely
recognized that it is important to take into account the effects of
the solvent-mediated long-range dynamic correlations between different
segments of the chain, known as hydrodynamic interactions
(HI). Although HI does not affect static properties, it seriously
alters the dynamic properties of semi-dilute and dilute solutions.
For instance, a recent study \cite{Frembgen-KesnerElcock09} of the
diffusion and folding of several model proteins has shown that
molecular simulations which neglect HI are incapable of reproducing
the expected experimental translational and rotational diffusion
coefficients of folded proteins, while simulations that include HI
predict the expected experimental values very well.  The study of the
folding process of small proteins has also revealed that inclusion of
HI hastens the folding process by at least a factor of two
\cite{Frembgen-KesnerElcock09,CieplakNiewieczerzal09}. The analogous
and simpler problem of the kinetics of homopolymer collapse both in
the absence and presence of HI has been studied via a variety of
different theoretical and numerical approaches
\cite{Lifshitzetal78,deGennes85,OstrovskyBaryam94,
  Kuznetsovetal95,TanakaMattice95,Byrneetal95,Buguinetal96,
  Grassberger97,Klushin98,HalperinGoldbart00,
  PolsonZuckermann00,ChangYethiraj01,Abramsetal02,
  PolsonZuckermann02,Kikuchietal02,Kikuchietal05,PolsonMoore05,
  LeeKapral06,Phametal08b}, and it is probably fair to say that a
reasonable degree of understanding has been obtained.

Although the homopolymer model has been widely used as a prototype to
understand the protein folding problem, it cannot capture all its
complexities. For instance, an important missing detail in a
homopolymer model is the formation of hydrophobic residues (H) or
amino acids in the core and polar residues (P) on the outer surface of
a collapsed globule. Moreover, it is believed that the kinetic ability
of natural proteins to fold results from the evolutionary selection of
the sequence of the molecule's amino acids, which should also result
in a unique native structure, i.~e. a collapsed state of vanishing or
at least low degeneracy in terms of its (free) energy. These are
aspects that cannot be captured by a homopolymer model; rather at
least two different types of interactions or ``letters'' are needed to
provide the possibility to encode a native structure.

For convenience as well as efficiency in numerical calculations, we
choose to use a simple two-letter code HP model to study the kinetics
of collapse of heteropolymers. Such models have been referred to as
prototeins \cite{Hayes98}. In this model, a chain is represented by a
binary sequence of H (hydrophobic) and P (polar) monomers. Although
the HP model has been widely used to study the protein folding problem
as well as heteropolymers in general, almost all the results are
obtained from predictions on discrete lattice models \cite{Hayes98,
  LauDill89, ChanDill96}, while direct simulation studies in the
continuum are scarce \cite{CookeWilliams03}. From such lattice models
one can conclude that only a small fraction of sequences folds
uniquely. Furthermore, not all geometric native structures can be
encoded in terms of a suitably adjusted sequence \cite{LauDill89,
  ChanDill96}.

For chains of reasonable length, the number of possible sequences and
conformations is too large for exact enumeration and hence statistical
sampling is required. In the present study, we restrict attention to
sequences that contain 50\% H monomers and 50\% P monomers; there
are both biological \cite{WhiteJacobs93} as well as theoretical
\cite{Dill85,CamachoThirumalai93b} arguments that this case should be
particularly relevant for protein folding. Previous theoretical and
simulational studies of heteropolymer collapse
\cite{Halperin91,Timoshenkoetal96,Villeneuveetal97,Timoshenkoetal98,
  KhokhlovKhalatur98,KhokhlovKhalatur99,Ganazzoli00,CookeWilliams03}
have shown that it typically comprises two stages, namely the rapid
formation of clusters or micelles along the chain, followed by coalescence, 
whose speed is rather sensitive to the details of the
sequence. Furthermore, the final geometry depends on these details as
well and is not always a compact spherical globule.

With the hope of gaining insight into the protein folding problem,
there has been a trend in recent years to design or engineer (in
theoretical or computer models) sequences that mimic biological
evolution such that they rapidly fold to a desired target native
structure
\cite{ShakhnovichGutin93a,ShakhnovichGutin93b,Shakhnovich94}. In
particular, Khokhlov and Khalatur \cite{KhokhlovKhalatur98,
  KhokhlovKhalatur99} have introduced such a method for a copolymer
chain composed of 50\% H and 50\% P monomers. It involves first
collapsing a homopolymer chain into its equilibrium compact state and
then marking half of the monomers that are closest to the centre of
mass as H type while the remaining monomers are marked as P type.
Copolymer chains with sequences generated via this method (also known
as ``protein-like'' copolymers) were found to fold much more rapidly
and the native conformations were more stable compared to random
copolymers and random block copolymers with the same average block
length of H or P polymers. This result was however obtained for a
lattice model without inclusion of HI. 

In the present study, we investigate the same question for an
off-lattice bead-spring model under both absence and presence of
HI. Furthermore, we investigate the influence of the length of H and P
blocks, i.~e. we compare the ``protein-like'' copolymers (PLC) not
only to random block copolymers with the same average block length
(RBC), but also to multi-block copolymers (MBC) with various different
block lengths. Finally, we attempt to identify the key features of an
HP sequence that govern the collapse kinetics.  In order to capture
realistic dynamics, we use Brownian Dynamics (BD) where HI is taken
into account via the Rotne-Prager-Yamakawa (RPY) tensor. To the best
of our knowledge, the combined effects of HI and chain sequence on
collapse kinetics within the framework of BD simulations has not been
reported so far in the literature.

In Sec.~\ref{sec:BS-model}, we provide the basic equations for the
bead-spring chain model and the BD simulation
method. Section~\ref{sec:Results} discusses our key findings, which
are summarised in Sec.~\ref{sec:Conclusion}.

\section{\label{sec:BS-model} Simulation method}

\subsection{\label{sec:Model} Model and equation of motion}

The macromolecule is represented by a bead-spring chain model of $N$
beads, which are connected by $N-1$ springs. Its configuration is
specified by the set of position vectors $\vecs{r}{\mu}$ ($\mu = 1,
2,\ldots N$). The dynamics is governed by the It\^{o} stochastic
differential equation \cite{OttingerBook96}
\begin{align}
\mathbf{r^*_{\mu}}(t^*+\Delta t^*) = \mathbf{r^*_{\mu}}(t^*) +
\fr{1}{4}\mathbf{D_{\mu\nu}} \cdot \cbr{{\mathbf{F}_\nu^\text{s}}^*
+{\mathbf{F}_\nu^{\text{int}}}^*}\Delta t^* 
+\fr{1}{\sqrt{2}}\mathbf{B}_{\mu\nu}\cdot\Delta\mathbf{W}_{\nu}
\quad \quad 
\nu = 1, 2,\ldots N .
\label{eq:SDE}
\end{align}
Here we use dimensionless units with $l_H=\sqrt{{k_\text{B}T}/{H}}$
and $\lambda_H=\zeta/4H$ as elementary length scale and time scale,
respectively \cite{PrabhakarPrakash04a,SuntharPrakash05}, where
$k_\text{B}$ is the Boltzmann constant, $T$ the temperature, $H$ 
the spring constant and $\zeta$ ($\zeta=6\pi\eta_s a$, where
$\eta_s$ is the solvent viscosity) the Stokes friction coefficient of
a spherical bead of radius $a$. Dimensionless quantities are
introduced via $\mathbf{r^*_{\mu}} = \mathbf{r_{\mu}} / l_H$ and $t^*
= t / \lambda_H$. $\mathbf{W_{\nu}}$ is a Wiener process,
$\mathbf{D_{\mu\nu}}$ is the diffusion tensor representing the effect
of the motion of a bead $\mu$ on another bead $\nu$ and is defined as
$\mathbf{D_{\mu\nu}} = \delta_{\mu\nu} \boldsymbol{\delta}
+\boldsymbol{\Omega}_{\mu\nu}$, where $\delta_{\mu\nu}$ is the
Kronecker delta, $\boldsymbol{\delta}$ is the unit tensor and
$\boldsymbol{\Omega}_{\mu\nu}$ is the hydrodynamic interaction
tensor. ${\mathbf{F}_{\nu}^\text{s}}^*$ is the dimensionless spring
force, ${\mathbf{F}_\nu^{\text{int}}}^*$ is the dimensionless
excluded-volume force and the components of $\mathbf{B_{\mu\nu}}$ are
related to the HI tensor such that $\mathbf{D_{\mu\nu}} =
\mathbf{B_{\mu \sigma}} \cdot \mathbf{B_{\nu \sigma}^\text{T}}$
\cite{OttingerBook96}. Throughout, the summation convention is implied
for repeated indices.

We use the regularised Rotne-Prager-Yamakawa (RPY) tensor to represent
HI \cite{RotnePrager69,Yamakawa71}; its form is
\begin{gather}
\mathbf{\Omega_{\mu \nu}} =
\mathbf{\Omega} ( \mathbf{r^*_{\mu}} - \mathbf{r^*_{\nu}} )
\end{gather}
with
\begin{gather}
\mathbf{\Omega}(\mathbf{r^*}) = 
\sbr{\Omega_1\boldsymbol{\delta}
+\Omega_2\fr{\mathbf{r^*}\mathbf{r^*}}{{r^*}^2}}
\end{gather}
and
\begin{equation}
\Omega_1 = \fr{3\sqrt{\pi}}{4} \fr{h^*}{r^*} 
\cbr{1+\fr{2\pi}{3}\fr{{h^*}^2}{{r^*}^2}}
\quad
\Omega_2 = \fr{3\sqrt{\pi}}{4} \fr{h^*}{r^*} \cbr{1-\fr{2\pi}{3}
\fr{{h^*}^2}{{r^*}^2}}
\quad 
\text{for} \,\,\, r^*\geq 2\sqrt{\pi}h^*
\end{equation}
\begin{equation}
\Omega_1 = 1- \fr{9}{32} \fr{r^*}{h^*\sqrt{\pi}} \quad
\Omega_2 = \fr{3}{32} \fr{r^*}{h^*\sqrt{\pi}} 
\quad 
\text{for} \,\,\, 0 < r^*\leq 2\sqrt{\pi}h^* ,
\end{equation}
where $h^*$ is the dimensionless bead radius in the bead-spring
model, defined as $h^* = a/(l_{H}\sqrt{\pi})$. Typical values of
$h^*$ lie between 0 and 0.5.

We have employed a two-letter code HP model for our copolymers, each
chain being composed of hydrophobic H and polar P beads. The polymer
is assumed to be in an aqueous solvent such that HP and PP
interactions are purely repulsive (i.~e.\ in good solvent), while HH
interactions are attractive (i.~e. in poor solvent). Similar to
previous work \cite{Phametal08b}, we use the dimensionless potentials
$V_\text{EV}$ for purely repulsive and $V_\text{attr}$ for attractive
interactions; they are defined as
\begin{align}
V_\text{EV}(r^*_{\mu\nu}) & = 
\left\{
\begin{aligned}
& 4 \fr{\epsilon_\text{LJ}}{k_\text{B}T}
\sbr{\fr{(\sigma/l_H)^{12}}{{r^*_{\mu\nu}}^{12}}
- \fr{(\sigma/l_H)^6}{{r^*_{\mu\nu}}^6}
+ \fr{1}{4}} 
& &
\text{for } r^*_{\mu\nu}\leq 2^{1/6}(\sigma/l_H),\\
& 0 
& & 
\text{for } r^*_{\mu\nu} > 2^{1/6} (\sigma/l_H) 
\end{aligned}
\right.\\
V_\text{attr}(r^*_{\mu\nu}) & = 
\left\{\begin{aligned}
& 4 \fr{\epsilon_\text{LJ}}{k_\text{B}T}
\sbr{\fr{(\sigma/l_H)^{12}}{{r^*_{\mu\nu}}^{12}}
- \fr{(\sigma/l_H)^6}{{r^*_{\mu\nu}}^6}
- c \, (R^*_c)}
& &
\text{for } r^*_{\mu\nu} \leq R^*_c,\\
& 0 
& &
\text{for } r^*_{\mu\nu} > R^*_c .
\end{aligned}
\right.
\label{eq:EV}
\end{align}
Here $\sigma$ and $\epsilon_\text{LJ}$ are the Lennard-Jones
parameters: $\epsilon^* = \epsilon_\text{LJ}/k_\text{B}T$ is the
dimensionless strength of the excluded volume interaction or quench
depth, which was used as a variable parameter; in most studies, we
chose the value $2.5$ (which is always assumed unless otherwise
stated). The parameter $\sigma/l_H$ denotes the corresponding
dimensionless length scale, which we set at $\sqrt{7}$. The
dimensionless distance from bead $\mu$ to bead $\nu$ is denoted by
$r^*_{\mu\nu}$. The cutoff radius of $V_\text{attr}$ is $R^*_c = 2.5
\, (\sigma/l_H)$ and the function $c \, (R^*_c)$ is chosen such that
the value of the potential is zero at the cutoff, i.e., 
$c \, (R^*_c)=[(\sigma/l_H)/R^*_c]^{12} - [(\sigma/l_H)/R^*_c]^6$.

The adjacent beads in the chain also interact via a finitely
extensible non-linear elastic (FENE) spring potential
\begin{equation}
V_\text{FENE} (r^*) = - \frac{1}{2}
\ln \left( 1 - \frac{r^{*2}}{Q_0^2} \right) ,
\end{equation}
where we choose $Q_0^* = 2 \, (\sigma/l_H) = 2 \sqrt{7}$ for the
dimensionless maximum stretchable length of a single spring.

The strength of HI is governed by the dimensionless Stokes radius
$h^*$. Motivated by previous work on FENE chains
\cite{Prabhakaretal04b,SuntharPrakash05}, we choose $h^* = 0.5
\sqrt{28 / 33} \approx 0.46$; this choice defines the HI strength as
$\tilde{h}^* = 0.5$ in units of the FENE equilibrium bond length
\cite{Prabhakaretal04b,SuntharPrakash05}. Our main guidance in
choosing $\tilde{h}^* = 0.5$ came from previous results on
homopolymers \cite{Phametal08b}, where this value led to a rapid
collapse. Similarly, we found in \Refcite{Phametal08b} that for
$\epsilon_\text{LJ}/k_\text{B}T = 2.5$ the homopolymer chain smoothly
folds into its final equilibrium compact stage without being trapped
in one of its intermediate metastable states.

For all simulations a time step size of $\Delta t^*=0.001$ was used,
yielding a solution with very small discretisation errors. For further
technical details on the BD algorithm, see
\Refcite{PrabhakarPrakash04a}.

The simulation procedure was carried out as a ``quenching'' 
experiment where the chain was started in a conformation corresponding to good solvent
conditions. The starting conformation was obtained by equilibrating a homopolymer where all
interactions are purely repulsive. This equilibration was done via a
BD run without HI for a period of $T_\text{eq} = 15 \,
\tau^*_{1,\text{R}}$, where $\tau^*_{1,\text{R}}=0.5\sin^{-2}(\pi/2N)$
is the longest dimensionless Rouse relaxation time, starting from a
random walk configuration. This procedure was repeated roughly 500
times in order to generate a sample of statistically independent
starting configurations which were stored for later use in the
quenching runs. At time $t = 0$, a sequence of H and P blocks was selected
(see below), and attractive interactions between HH pairs were turned
on. From then on, the collapse process was monitored.

\subsection{\label{sec:Sequence} Chain sequence construction}

We have carried out simulations for two different chain lengths of $N
= 64$ and $128$ beads at a fixed H:P ratio of $1:1$ (i.~e.
$N_\text{H} = N_\text{P} = N/2$). We have studied three different
families of copolymer chains with sequence types that were either
inherited from the parent globule, regular or probabilistic. 

The first sequence type is the protein-like copolymer or PLC, where the sequence 
has been generated by the following process. A homopolymer is picked 
from the set of 500 stored initial conformations, and then collapsed to a compact globule by making 
all the beads attractive. This is followed by marking the interior as 
hydrophobic H monomers and the exterior as hydrophilic or polar P
ones. Typically, this procedure leads to a sequence of H and P blocks which look at
least partially disordered when viewed along the chain backbone. The PLC sequence 
generated in this way is then used to colour a homopolymer from the existing pool of good-solvent conformations, which is different from the one used to initiate the collapse process. 
This procedure is repeated 500 times, always ensuring that 
the homopolymer conformation that is decorated with the PLC sequence obtained at 
the end of the collapse and colouring process, is different from the starting homopolymer conformation.
In this way, we ensure that we have a set of 500 PLC chains that differ from each other in sequence and in 
conformation. During the quench experiment, each of 
these PLC chains is then collapsed again, leading to 500 independent quench runs per parameter set.  

The second family of copolymers is the regular or alternating
multi-block copolymer (MBC) with block length $L$ (number of monomers
in a contiguous H or P block). Hence the sequences were of the form
$\left( \text{H}_L \text{P}_L \right)_n$ where $n = N/(2L)$ is the
number of block pairs. No randomness is involved in this architecture;
therefore again the sample size is approximately 500 independent
quench runs per parameter set.

The last family is known as the random-block copolymer (RBC) in which
each block length $L$ is an independent random variable sampled from a
Poisson distribution
\begin{equation}
f(L) = \frac{\lambda^L}{L!} \exp \left( - \lambda \right) ,
\end{equation}
where $\lambda = \avbr{L}$ is the average block length. The Poisson
random variables were generated from a built-in function provided in
MATLAB. For each of the $\sim 500$ stored starting conformations we
generated one new sequence, such that again we have a total of $\sim
500$ independent quench runs per parameter set. We imposed a strict
$1:1$ constraint on the ratio of the number of monomers
(H:P). Therefore, the H and P blocks were treated separately. Since
the Poissonian random numbers usually do not add up to precisely
$N/2$, we simply replaced the last random number with the remaining
block.

\subsection{\label{sec:Observables} Observables}

We monitor the time dependence of various observable quantities given
below, which can be used to characterise the collapse
kinetics. Important observables are the mean square radius of gyration
\begin{equation}
\avbr{{R_\text{g}^*}^2} = 
\frac{1}{2 N^2} \sum_{\mu \nu} \avbr{{r^*}^2_{\mu \nu}}
\end{equation}
and the internal energy
\begin{equation}
U = \sum_{\mu < \nu} \left<
V_\text{attr}(r^*_{\mu \nu}) c_\mu c_\nu
+ V_\text{EV} (r^*_{\mu \nu})
\left[ 1 - c_\mu c_\nu \right] \right> ,
\end{equation}
where $c_\mu$ is unity for an H monomer and zero otherwise.

The instantaneous shape of a polymer chain can be determined from its
gyration tensor, $\mathbf{G}$, whose Cartesian components are given by
\begin{equation}
G_{ij} = \frac{1}{N} \sum_{\mu}
\left( r_{\mu i} - r_{\text{cm} i} \right)
\left( r_{\mu j} - r_{\text{cm} j} \right) ,
\end{equation}
where the cm subscript refers to the chain's centre of mass. For each
conformation, we determine the three eigenvalues $\lambda^2_1$,
$\lambda^2_2$, $\lambda^2_3$ of $\mathbf{G}$, using the convention
$\lambda^2_1 \ge \lambda^2_2 \ge \lambda^2_3$. The ratios of these
numbers give an indication of the deviation of the molecule's shape
from a sphere.

The total collapse time $\tau$ is defined as the time needed for the
radius of gyration to reach 99\% of its total change in size during
the transition period in which the chain is transformed from the
initial coil to the final state, i.~e.
\begin{equation}
R^*_\text{g}(\tau) - R^*_\text{g, eq}
= \fr{1}{100} (R^*_\text{g}(0) - R^*_\text{g,eq})
\end{equation}
where $R^*_\text{g,eq}=\sqrt{\avbr{{R_\text{g}^*}^2}_\text{eq}}$ is
the root mean square equilibrium dimensionless radius of gyration in
the collapsed state (defined as the value obtained at the time at which the run was stopped). 

We also monitor the time dependence of the average cluster
size, defined as
\begin{equation}
\avbr{S_n(t^*)} = \fr{\sum_s sn(s)}{\sum_s n(s)},
\end{equation}
using the algorithm of \Refcite{Sevicketal88}, 
where $n(s)$ denotes the number of clusters of size $s$. Our previous investigation on
homopolymers \cite{Phametal08b} showed that this definition gives a
more consistent picture of the dynamics than other possible
definitions involving higher-order moments. Only H-type beads were
used to define a cluster. Two H monomers are considered to be part of
the same cluster if they (i) are not nearest neighbours along the
backbone of the chain, and (ii) have an interparticle distance that
does not exceed a certain value $D$, which we have set (in our units)
to $D^* = 1.2 \times 2^{1/6} (\sigma / l_H)$, following previous
studies \cite{Byrneetal95,CookeWilliams03,Phametal08b}.

It is appropriate to note that all averages reported here are estimated by carrying out sample averages across the 500 independent runs, and the reported error is the estimated error in these averages. Except in the case of MBC chains, where the sequence is identical across all starting conformations, our data therefore involves double sampling over sequences and starting conformations. In order to avoid overcrowding in the plots, the statistical error bars are only included for a few selected points. In cases where the error bars are smaller than the symbol size, they are omitted.

\section{\label{sec:Results} Results and discussion}
%
\begin{figure}[tbp!]
\centering \subfigure[] {
   {\includegraphics[width=0.8\textwidth]{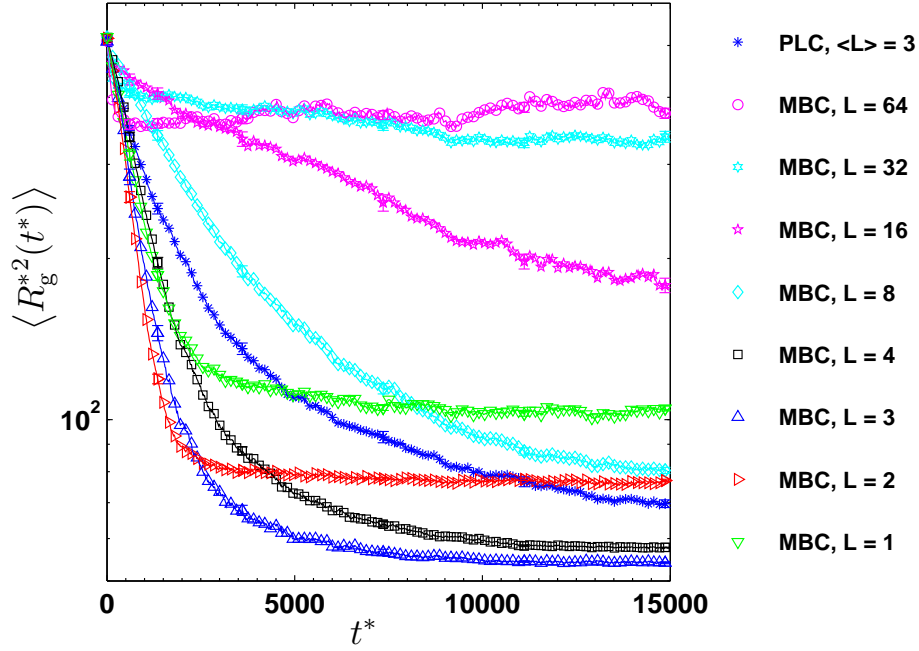} }
}  \subfigure[] {
    {\includegraphics[width=0.8\textwidth]{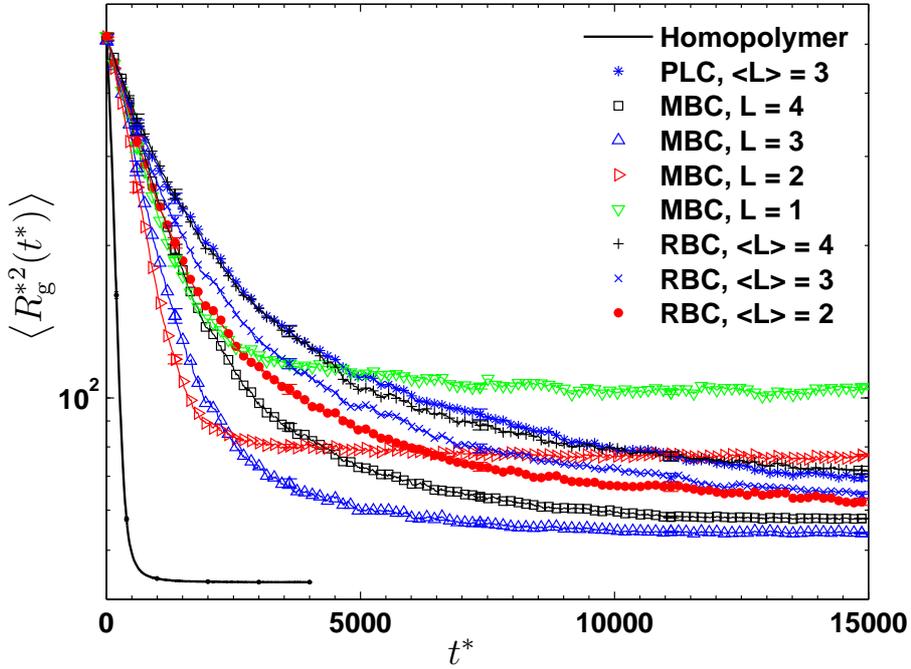}}
}
\vskip-40pt
\caption{Variation of the mean square radius of gyration with time for
  $N=128$ chains, in the presence of HI, for (a) the PLC chain and MBC
  chains with various values of block length $L$, and (b) all three
  types of copolymers with values of average block length
  $\avbr{L}\leq 4$.  }
\label{fig:RgsqvstHI}
\vskip-20pt
\end{figure}
\subsection{\label{sec:Size} Chain size}

Figure~\ref{fig:RgsqvstHI} shows the effects of the block length $L$
for the protein-like, regular multi-block and random block copolymer
chains on the evolution of the mean square radius of gyration for
$N=128$, in the presence of HI. Obviously, the average block length
plays an important role in controlling the dynamics of the process as
well as the final equilibrium size of the copolymer chains. For the
PLC chains in our model, the average block length $\avbr{L}$, for both
$N=64$ and $N=128$ bead chains, is approximately 3, which is very
close to the value reported for the analogous lattice model
\cite{KhokhlovKhalatur98, KhokhlovKhalatur99} ($\avbr{L}=3.173$).

We first discuss the effect of block length on the behaviour of the MBC
chain. The simplest type of MBC is the diblock copolymer with
$L=N/2$. It can be seen from Fig.~\ref{fig:RgsqvstHI} (a) that this
diblock chain collapses to its equilibrium state faster than any of
the other copolymer chains present in the plot. This is always true
for such a copolymer because one of its ends (i.~e. the end with the H
block) behaves like a homopolymer chain in a poor solvent, which
rapidly folds into its final equilibrium globular state, while the
other end behaves like a homopolymer in a good solvent, remaining as a
swollen coil. Due to the large size of this swollen part, the chain
cannot form a compact equilibrium structure and consequently always
has a relatively large final equilibrium size. 

By reducing the block size to $L=32$, such that one has 4 blocks, one
can see two time regimes of collapse. The first regime represents the
early stage of collapse, where each H block of size $L$ rapidly folds
into a single cluster. Thus at the end of this stage, the chain
consists of $N/(2L)$ clusters or pearls separated by strings of P
block chains. Since the folding of the H block is entirely
homopolymer-like and is unaffected by the P type monomers, the time
taken for each of these H block chains to completely fold into a small
cluster is quite small, and it becomes systematically smaller for
decreasing block size, such that it becomes hardly observable for
shorter blocks.

The second regime represents the growth or coalescence of these
intrachain clusters into a single large cluster. These clusters are
separated from each other by a string of P block, which has the same
size as that of the original H block. If the separation between the
two intrachain clusters is larger than the range of the attractive
interaction (or, equivalently, if the P block length is large enough),
then these intrachain clusters go through a diffusion-like process. It
is only when they interact with each other, i.~e. when their
separation distance comes within the range of attractive interaction,
that they coalesce. Thus the growth of the average cluster size is
quite gradual in this case. However, if the P block chain length is
sufficiently small, such that two intrachain clusters are within the
range of their attractive interaction, then the two clusters quickly
coalesce to form a larger cluster and diffusion plays little or no
role. Moreover, the average cluster size will also grow much more
rapidly, or, equivalently, there is a rapid reduction in chain
size. Since the diffusion time increases with the size of the
intervening P block \cite{CookeWilliams03}, it is expected that the
coalescence time is shorter for smaller P block sizes and consequently
the collapse is much more rapid compared to large P block sizes.
Furthermore, a large P block size leads to a less compact final
collapsed size because blocks of P beads are always repelled from the
core of H beads, and they form dangling legs extruding away from the
core. The results for $L=32$ to $L=8$ for MBC chains in
Fig.~\ref{fig:RgsqvstHI} (a) also confirm these expectations.

Note that for all MBC chains with $L \geq 8$, the final equilibrium
size is still larger than that of the PLC chain. However, when $L=4$,
the MBC chain reaches its equilibrium native state faster, and the
final equilibrium size is smaller than that of a PLC chain. Reducing
the block size of the MBC chain to $L=3$ further reduces the final
size as well as the total collapse time. In fact, the data shows that
the MBC chain with $L=3$ produces the lowest final equilibrium size
compared to all other types of copolymer chains for the entire range
of $L$ and $\avbr{L}$ values that have been investigated. The total
collapse time and the mean square radius of gyration for all three
types of copolymers are listed in Table~\ref{tab:exponents}.

\begin{table}[!t]
\begin{center}
\begin{tabular}{l|ccccc}
\hline\hline 
Type & $\avbr{L}$ & $\alpha$ & $z$ & $\tau$ & 
$\avbr{R_\text{g}^2}_\text{eq}$ \\
\hline 
PLC & 3 & 0.945 $\pm$ 0.021 & 0.510 $\pm$ 0.009 & 5732 $\pm$ 146 & 
69.762 $\pm$ 1.151 \\
MBC & 64 & 0.999 $\pm$ 0.012 & 0.920 $\pm$ 0.044 & 2044 $\pm$ 185 & 
369.034 $\pm$ 8.605 \\
MBC & 32 & 0.978 $\pm$ 0.022 & 0.158 $\pm$ 0.003 & N/A & N/A \\
MBC & 16 & 0.842 $\pm$ 0.043 & 0.441 $\pm$ 0.005 & N/A & N/A \\
MBC &  8 & 0.616 $\pm$ 0.020 & 0.646 $\pm$ 0.004 & 6670 $\pm$ 139 & 
79.819 $\pm$ 0.989 \\
MBC &  4 & 1.033 $\pm$ 0.026 & 0.819 $\pm$ 0.017 & 4248 $\pm$ 101 & 
57.616 $\pm$ 0.064 \\
MBC &  3 & 1.316 $\pm$ 0.088 & 0.858 $\pm$ 0.018 & 3304 $\pm$ 118 & 
54.287 $\pm$ 0.239 \\
MBC &  2 & 1.287 $\pm$ 0.029 & 1.008 $\pm$ 0.025 & 1967 $\pm$ 53 & 
77.051 $\pm$ 0.385 \\
MBC &  1 & 0.929 $\pm$ 0.038 & 0.857 $\pm$ 0.018 & 2771 $\pm$ 103 & 
103.906 $\pm$ 1.134 \\
RBC &  4 & 0.987 $\pm$ 0.028 & 0.631 $\pm$ 0.008 & 5447 $\pm$ 138 & 
71.511 $\pm$ 1.278 \\
RBC &  3 & 1.071 $\pm$ 0.025 & 0.575 $\pm$ 0.006 & 5496 $\pm$ 141 & 
64.513 $\pm$ 1.006 \\
RBC &  2 & 1.045 $\pm$ 0.037 & 0.595 $\pm$ 0.009 & 4948 $\pm$ 132 &
62.839 $\pm$ 0.985 \\
\hline\hline
\end{tabular}
\end{center}
\caption{Values of the exponents for the early stages of collapse and
  the growth of the number average cluster size (i.~e. $\alpha$ and $z$
  in $\avbr{R^2_\text{g}(0)}-\avbr{R^2_\text{g}(t)}\sim t^{\alpha}$
  and $\avbr{S_n}\sim t^z$) for all types of copolymers with $N=128$
  at various values of average block lengths $\avbr{L}$, in the
  presence of HI. Values of the total collapse time $\tau$ and the
  equilibrium mean square radius of gyration
  $\avbr{R^2_\text{g}}_\text{eq}$ are listed as well. Note that PLC,
  MBC and RBC denote protein-like, multi-block and random-block
  copolymers, respectively.}
\label{tab:exponents}
\vskip-20pt
\end{table}

Further reduction in the block size to $L = 2$ leads to a slightly
faster collapse (as evidenced by the value of $\tau$ in
Table~\ref{tab:exponents}), but it also increases the final chain
equilibrium size. Note that for every $L$ monomers of H type that fuse
with another H cluster, $L$ monomers of P type are brought into close
proximity, due to the connectivity along the chain. As a result, a
repulsive force builds up as the coalescence process takes place. For
the $L = 2$ chain, the energy gained from coalescence cannot overcome
this repulsion, and therefore the chain does not form a single cluster
of H monomers, i.~e. it cannot fold into a spherical compact globule.
The $L = 1$ case conforms to this behaviour, with the chain collapsing
slower, and the final equilibrium size being even larger than for
$L=2$.

\begin{figure}[tbp]
\centerline{\includegraphics[width=13cm,height=!]{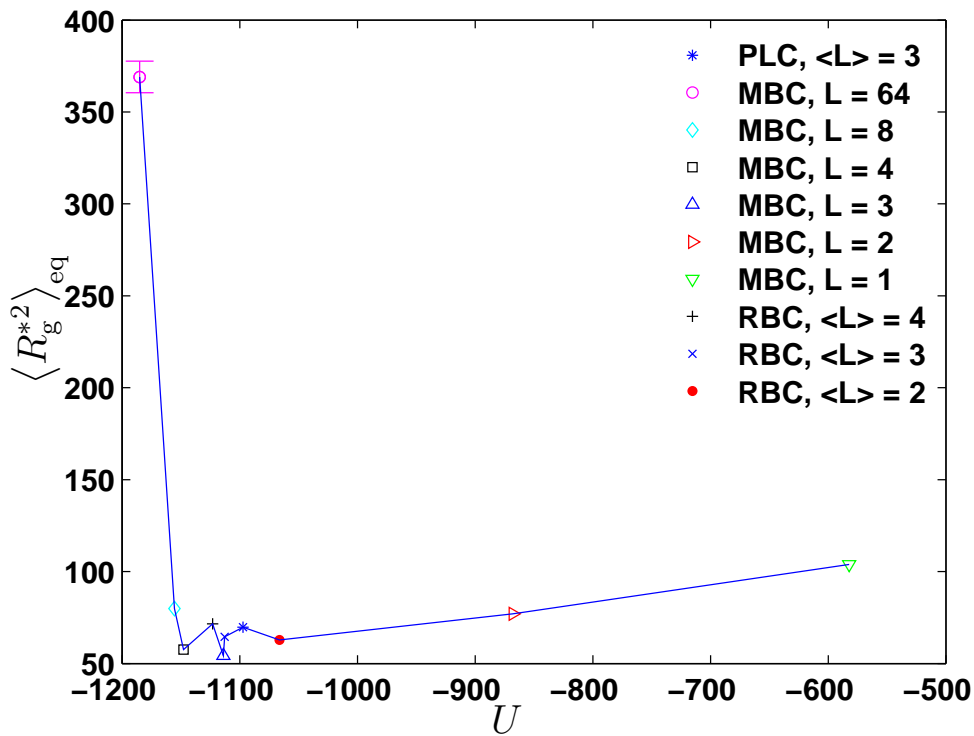}}
\vskip-40pt  
\caption{Coordinate pairs ($U,\avbr{{R_\text{g}^*}^2}_\text{eq}$) of
  the internal energy and $\avbr{{R_\text{g}^*}^2}_\text{eq}$,
  respectively, in the final collapsed state for all copolymer chains
  with $N=128$, in the presence of HI.}
\label{fig:RgsqeqvsEVHI}
\vskip-20pt
\end{figure}

The large repulsive force built up due to the overcrowded presence of
P type monomers for MBC chains with $L\leq 2$ can also be observed via
the high value of the internal energy $U$ in
Fig.~\ref{fig:RgsqeqvsEVHI}. Note that the internal energy for MBC
chains with $L=32$ and 16 is not reported here because the chains with
these block sizes had not yet reached their equilibrium state, which
is also subject to large fluctuations. This figure also clearly
indicates that a chain which has the lowest equilibrium size does not
necessarily have the lowest internal energy. Moreover, a chain with
the lowest internal energy may not fold into the most compact final
structure. This finding is consistent with the results of
\Refcite{CookeWilliams03}, where BD simulations without HI were
performed.


From Fig.~\ref{fig:RgsqvstHI} (b), we can see that the PLC chain and
the RBC chain with $\avbr{L} = 4$ are almost identical in terms of
their final sizes and their time behaviour of $R_\text{g}$. However,
reducing the average block size of the RBC chain to $\avbr{L} = 3$ and
$\avbr{L} = 2$ systematically decreases both the collapse time and the
equilibrium size even further. This finding is at variance with the
results of \Refcite{KhokhlovKhalatur99}, where the PLC chain was
found to collapse more rapidly and into a more compact structure than
the RBC chain with the same $\avbr{L}$. It is not quite clear what the
origin of this discrepancy is; most likely it is the fact that our
model is a bead--spring model in the continuum, while
\Refcite{KhokhlovKhalatur99} studied a lattice model, which results
in different local packing geometries, which are certainly very
important for this problem. Another possible source could be different
interaction parameters. The fact that we study a system with HI and
\Refcite{KhokhlovKhalatur99} one without can however \emph{not} be
used as an explanation; as Fig.~\ref{fig:Rgsqvst} shows, we find the
same behaviour also in the absence of HI (see below). For the protein
folding problem this result indicates that the relation between the
sequence and the thus-encoded collapse kinetics and native state is
probably more intricate than the simple and appealing picture that the
labeling procedure of \Refcite{KhokhlovKhalatur99} suggests.


\begin{figure}[tbp]
\centerline{\includegraphics[width=13cm,height=!]{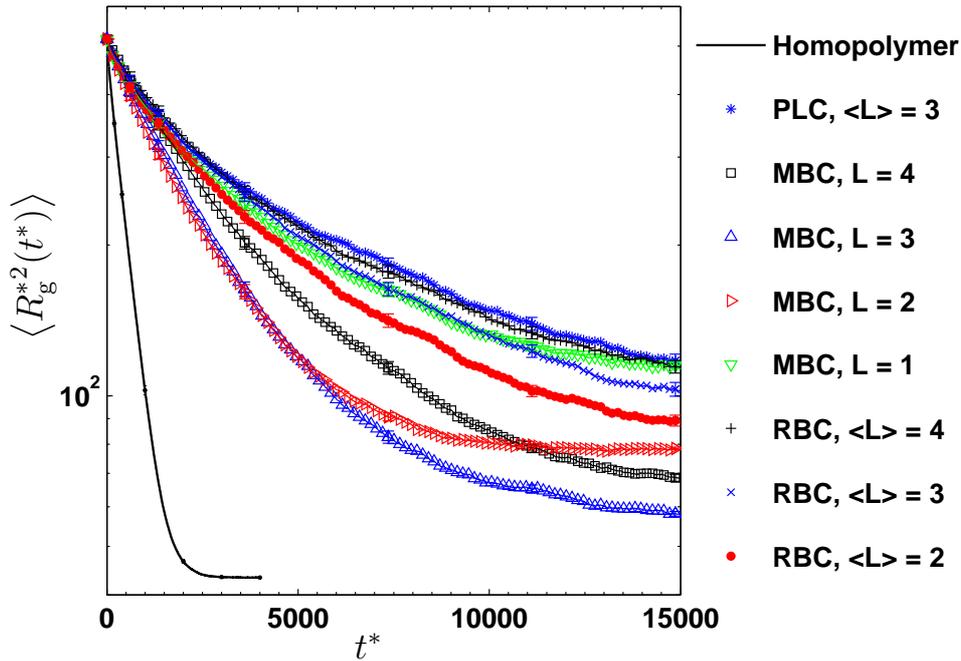}}
\vskip-40pt  
\caption{Variation of the mean square radius of gyration with time for
  all three types of copolymer chains with $N=128$, for various values
  of the average block length $\avbr{L}$, in the absence of HI.}
\label{fig:Rgsqvst}
\vskip-20pt
\end{figure}

We fitted a power law
\begin{equation}
\avbr{R^2_\text{g}(t)} = \avbr{R^2_\text{g}(0)} - A t^\alpha
\end{equation}
to the initial decay of the mean square gyration radius, and the
resulting exponents $\alpha$ are listed in Table~\ref{tab:exponents}.
For a homopolymer chain at the same quench depth,
\Refcite{Phametal08b} had obtained a value of $\alpha = 1.05 \pm
0.01$ in the presence of HI. Interestingly, it can be seen from
Table~\ref{tab:exponents} that there exist some copolymer chains which
have a larger exponent for this early stage compared to that of a
homopolymer chain. Similar results have also been observed in
\Refcite{CookeWilliams03} for copolymer chains in the absence of HI.


%
\begin{figure}[tbp!]
\centering \subfigure[] {
   {\includegraphics[width=0.8\textwidth]{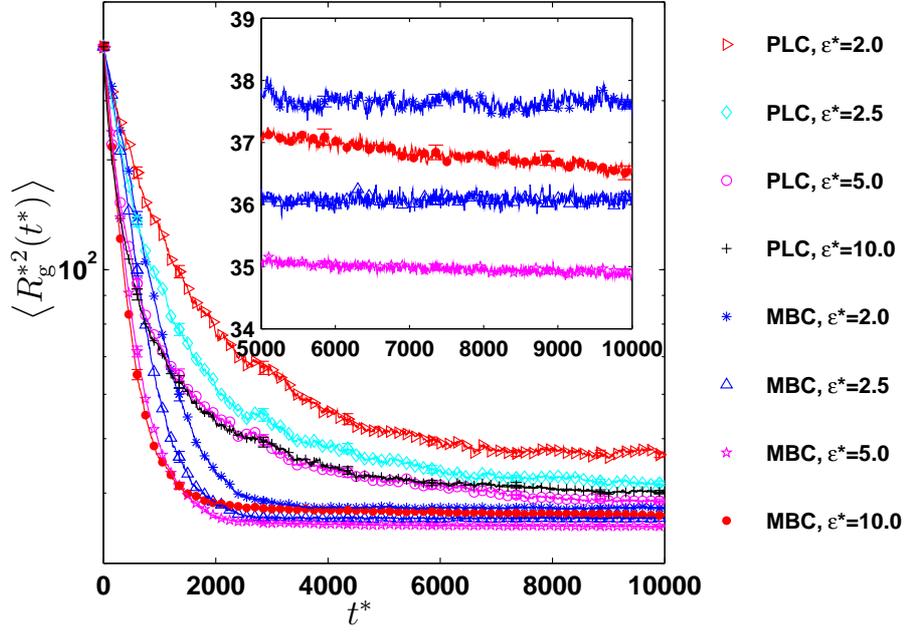} }
}  \subfigure[] {
    {\includegraphics[width=0.8\textwidth]{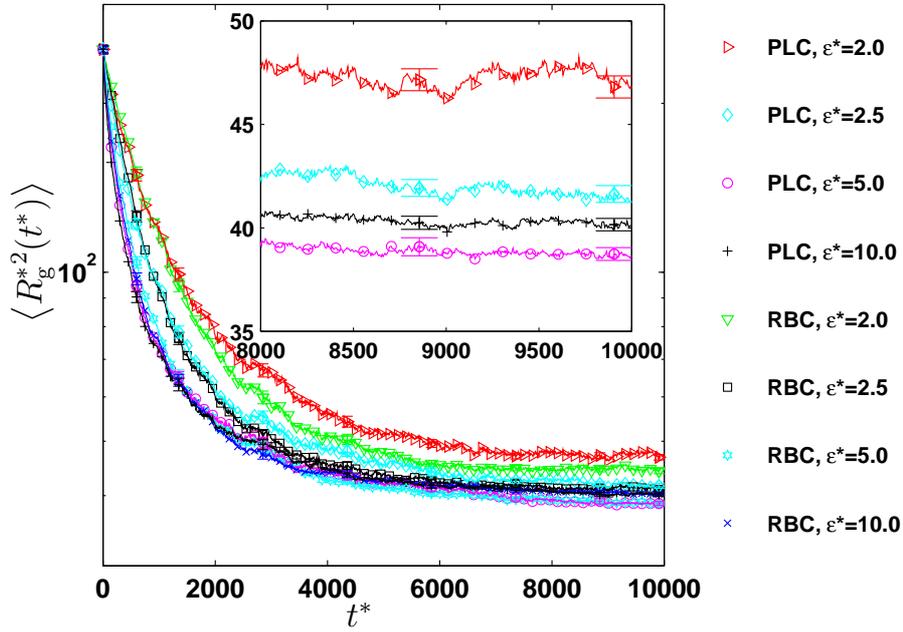}}
}
\vskip-40pt  
\caption{Variation of the mean square radius of gyration with time for
  chains with $N=64$, in the presence of HI, for (a) PLC and MBC
  chains with block length of $L=3$, and (b) PLC and RBC chains with
  an average block length of $\avbr{L}=3$. Here, $\epsilon^*$ denotes
  the value of the quench depth or $\epsilon_\text{LJ}/k_\text{B}T$.
  Insets: Variation of the mean square radius of gyration with time
  for chains with $N=64$ and $L=3$ or $\avbr{L}=3$ at various quench
  depths $\epsilon^*$, for (a) MBC chain, and (b) PLC chain.}
\label{fig:N64RgsqvstEVHI}
\end{figure}

\subsection{\label{sec:InfluenceHI} Influence of HI}

In order to study the influence of HI, we have also carried out
simulations without HI, for the case $N = 128$, and the results are
shown in Fig.~\ref{fig:Rgsqvst}. We find that in general the collapse
dynamics is substantially slowed down, compared to the HI case, which
corresponds to the observation for homopolymers that the presence of
HI (Zimm model) leads to systematically faster dynamics than in the case when 
HI is absent (Rouse model). In most cases, the chains had not even
relaxed fully into equilibrium at the time at which the HI runs (and
also the non-HI runs) were stopped. For the one case in which
equilibrium was reached, we find the same value for $\left<
R_\text{g}^2 \right>_\text{eq}$ as in the case with HI, as it must
be. The qualitative results concerning the influence of the chain
architecture on the collapse kinetics remain however completely
unchanged: Again, the PLC chain corresponds to the $\avbr{L} = 4$ RBC,
but the RBCs with shorter block length collapse even faster.

The observed acceleration of the collapse dynamics as a result of HI
is consistent with our previous findings on homopolymer collapse
\cite{Phametal08b}, where it was shown that for $N = 128$ the
inclusion of HI reduces the total collapse time by at least a factor
of two. A complete comparison between the total collapse times for
cases with and without HI is not reported in this work, due to the
slow convergence to equilibrium in the absence of HI. The only case
where equilibrium was reached (MBC, $L = 2$) seems to indicate that
the speedup is even larger in the present case ($\tau = 5233$ without
HI, $\tau = 1967$ with HI). Since hydrodynamics is a large-scale
phenomenon, one expects the effect of HI to be the more dramatic the
more extended the objects are. Due to the uncollapsed P strings, the
chains stay systematically longer in an extended state than in the
homopolymer case, and this is probably the reason why HI has an even
stronger effect.

\subsection{\label{sec:InfluenceQuench} Influence of quench depth}

We have simulated $N = 64$ chains at various values of the quench
depth $\epsilon^*$, in order to study the influence of this parameter,
while keeping the block length (or average block length) at $L = 3$
(or $\avbr{L} = 3$). Figure~\ref{fig:N64RgsqvstEVHI} shows the results
for the gyration radius. For all the quench depths investigated, it
can be seen that the MBC chains collapse the fastest and have the most
compact equilibrium size, followed by the RBC chains. Out of all three
types of copolymer chains, the PLC chain collapses the slowest and the
final equilibrium size is the least compact. Generally, one would
expect that the radius of gyration for a chain at deep quench should
be smaller than that of a chain with lower quench values because the
stronger attractive interaction would cause the chain to squeeze into
a tighter globule. The above figure shows that a chain smoothly folds
into its final compact state for low quench depth. However, at very
large quench depth (i.~e. $\epsilon_\text{LJ}/k_\text{B}T=10$), a
chain gets trapped in a metastable state and stays there for a long
time rather than approach its final minimum energy state. The inset of
Fig.~\ref{fig:N64RgsqvstEVHI} (a) shows this trapping behaviour at
very large quench depths much more clearly for an MBC chain. This
inset reveals that, while MBC chains with
$\epsilon_\text{LJ}/k_\text{B}T\leq 5$ have fully reached their
equilibrium compact state, chains with
$\epsilon_\text{LJ}/k_\text{B}T=10$ are still gradually approaching
the equilibrium state. A similar result is also observed for other
types of copolymer chains (see Fig.~\ref{fig:N64RgsqvstEVHI} (b)).  A
similar behaviour has been observed by various other authors
\cite{ChangYethiraj01, Kikuchietal05, Phametal08b, Zivetal08}. In
\Refcite{Phametal08b} it was found that a homopolymer chain will get
trapped at $\epsilon_\text{LJ}/k_\text{B}T=5$, while our results
indicate that a copolymer chain still smoothly folds at this quench
depth. This result seems to indicate that the presence of P type
monomers in the chain prevents it from being trapped in a local well
and smooths out the energy landscape for the folding process of
copolymers. This in turn pushes the value of quench depth where
trapping occurs to a much higher value.


\begin{figure}[tbp]
\centerline{\includegraphics[clip,width=0.95\textwidth]{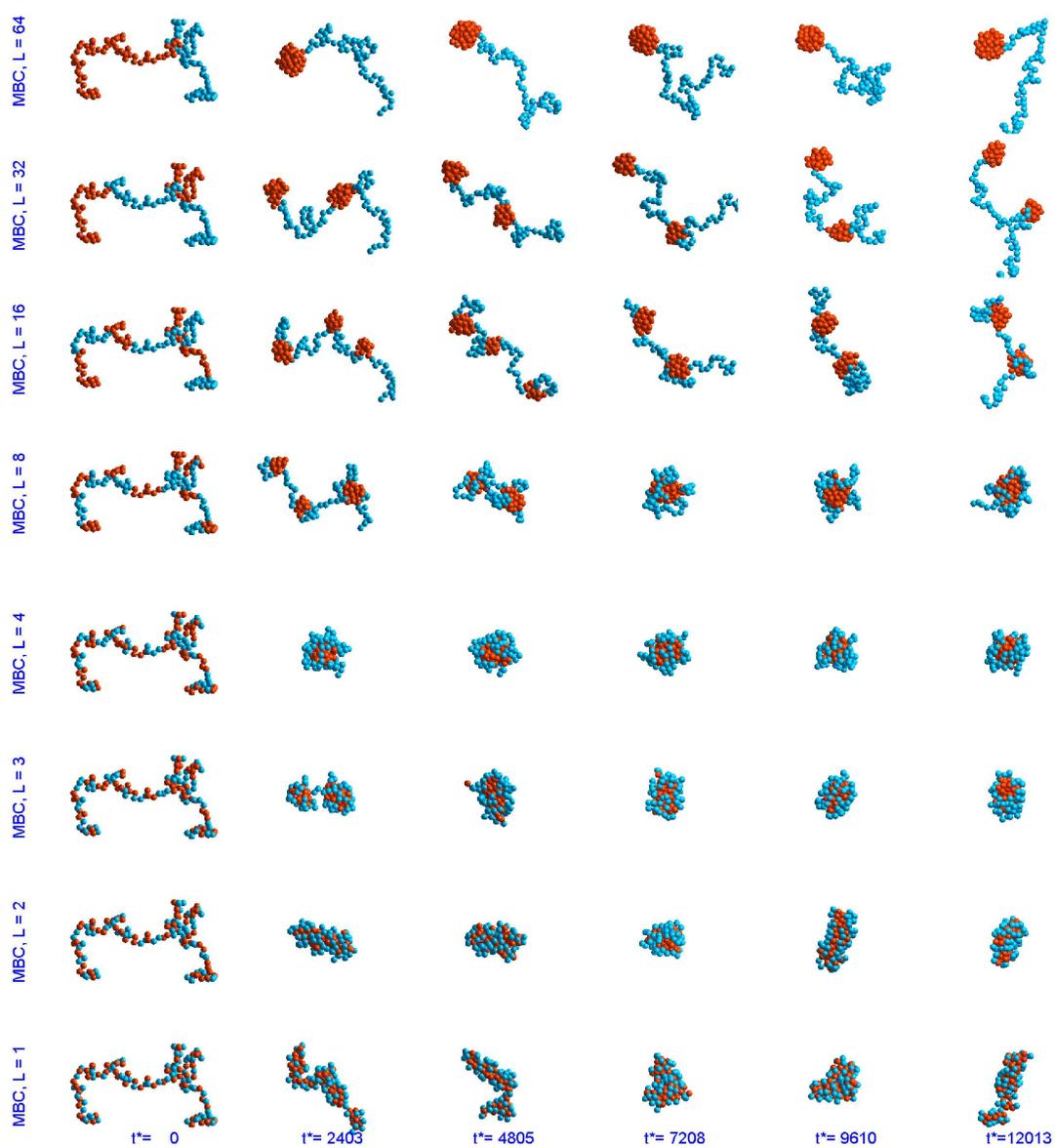}}
\vskip-40pt  
\caption{Snapshots of different types of collapsing regular
  multi-block copolymers (MBC) chain with $N=128$, in the presence of
  HI. }
\label{fig:MBCpath}
\end{figure}


\begin{figure}[tbp]
\centerline{\includegraphics[width=0.95\textwidth]{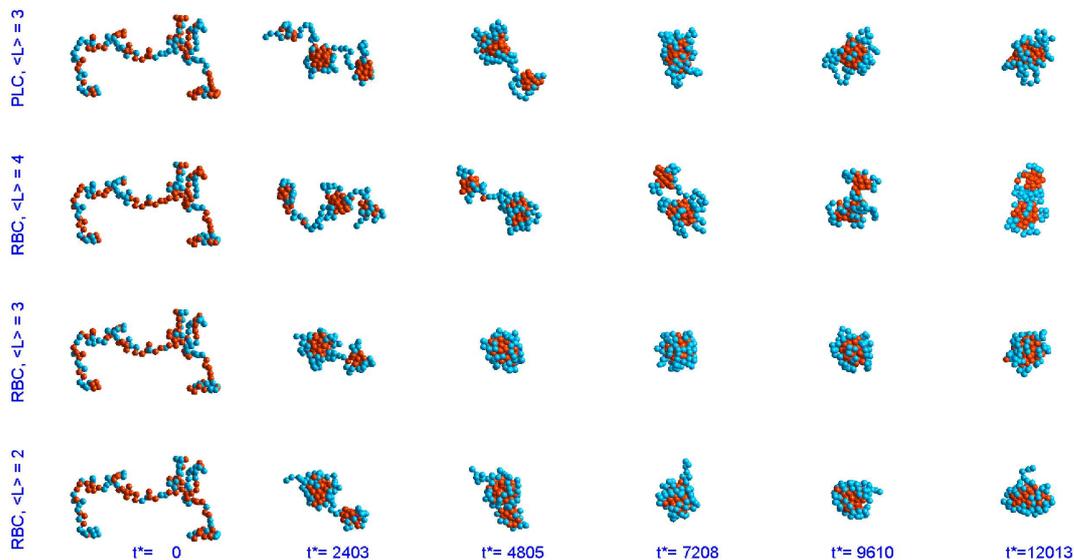}}
\vskip-40pt  
\caption{Snapshots of different types of collapsing protein-like
  copolymers (PLC) and random-block copolymers (RBC) chains with
  $N=128$, in the presence of HI.}
\label{fig:PLCRBCpath}
\vskip-20pt
\end{figure}


\begin{figure}[!tbp]
\centerline{\includegraphics[width=12cm,height=!]{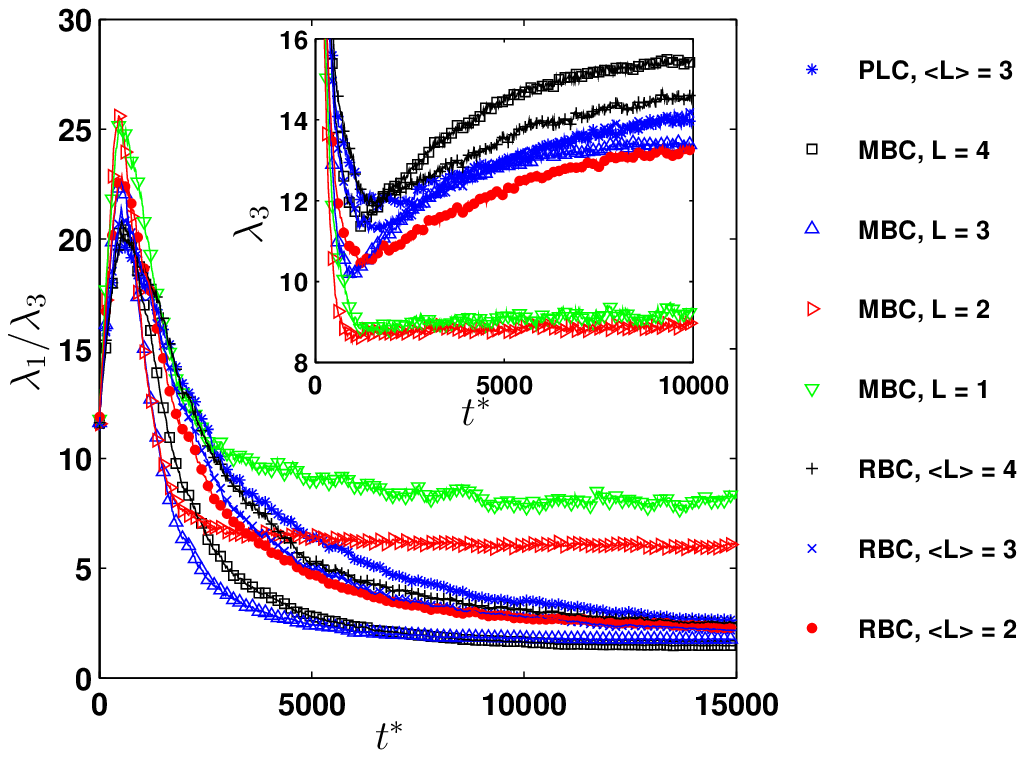}}
\vskip-40pt  
\caption{Variation of the ratio of the largest eigenvalue to the
  smallest eigenvalue of the gyration tensor with time for all three
  types of copolymer chains with $N=128$, for various values of the
  average block length $\avbr{L}$, in the presence of HI. Inset: The
  corresponding variation of the smallest eigenvalue with time.
  Here, the symbol $\lambda_i$ is actually an abbreviation for
  $\avbr{\lambda_i^2}^{1/2}$.}
\label{fig:Eig1by3vstHI}
\vskip-20pt
\end{figure}

\subsection{\label{sec:Snapshots} Snapshots}

Snapshots of the typical kinetic pathways for a collapsing chain of
length $N=128$ for MBC chains are shown in Fig.~\ref{fig:MBCpath}, and
for PLC and RBC chains in Fig.~\ref{fig:PLCRBCpath}. For some  selected cases, 
we show the time evolution in the movie files S1.avi and S2.avi of the Supporting Information.
It can be seen that there exist at least two distinct stages
for the kinetics of collapse for these copolymers. It is to be noted
that almost all of the copolymer chains with sequences that fold into
a spherical compact structure have a three-stage mechanism. The early
stage involves the rapid formation of localised globules along the
chain. Following this stage, these globules then coalesce together by fusing
with other nearby blobs to form a dumbbell or two pearls separated by
linear chain. Finally, the pearls combine to form a sausage which
slowly rearranges itself into a compact state. However, for copolymer
chains with sequences consisting of short P blocks that do not fold
into a spherical compact structure (for instance, MBC chains with $L
\leq 2$), only the first two distinct stages are observed and the
final compaction stage apparently does not occur. Furthermore,
copolymer chains with sequences consisting of large P blocks, such as
MBC chains with $L = 32$ and 16, seem to acquire another distinct
stage after the early rapid collapse stage, known as the ``diffusion''
or ``plateau'' regime. During this diffusion stage, the number and
size of the intrachain clusters remains unchanged due to the large
separation between the H blocks. Similar qualitative features of the
collapse pathways have also been observed by other authors
\cite{CookeWilliams03,Dasmahapatraetal06}. 

\subsection{\label{sec:aspher} Asphericity}

Apart from the snapshots, the asphericity of the chain can also be
observed from the eigenvalues of the gyration tensor.
Figure~\ref{fig:Eig1by3vstHI} shows the time evolution of the ratio of
the largest eigenvalue to the smallest eigenvalue for all three types
of copolymer chains with $N = 128$, for various values of the average
block length $\avbr{L}$, in the presence of HI. Interestingly, the data
in Fig.~\ref{fig:Eig1by3vstHI} clearly shows that this ratio of
eigenvalues behaves non-monotonically with time. A similar result is
also observed for the ratio of the intermediate eigenvalue to the
smallest eigenvalue $\lambda_2/\lambda_3$. The non-monotonic behaviour
is mainly due to the non-monotonic nature of the smallest eigenvalue
$\lambda_3$, as can be seen in the inset of
Fig.~\ref{fig:Eig1by3vstHI}. This behaviour can be explained by the
fact that during the collapse of the chain, there is a sudden change
in the shape of the chain from an initial ``ellipsoid'' (note that a
typical self-avoiding walk is highly anisotropic) to a pearl-necklace
and finally (typically) to a sphere. During this shape transformation,
the eigenvalue for the smallest principal axis is first reduced to the
lowest value, at which the thinnest pearl-necklace chain is
formed. This principal axis then grows in width when the
pearl-necklace chain starts to transform by absorbing nearby pearls or
clusters, and when the bridging strings of monomers form a final
spherical structure. Note that a similar result has also been observed
for a homopolymer chain~\cite{Phametal08b}.

\begin{figure}[!tbp]
\centerline{\includegraphics[width=!,height=14cm]{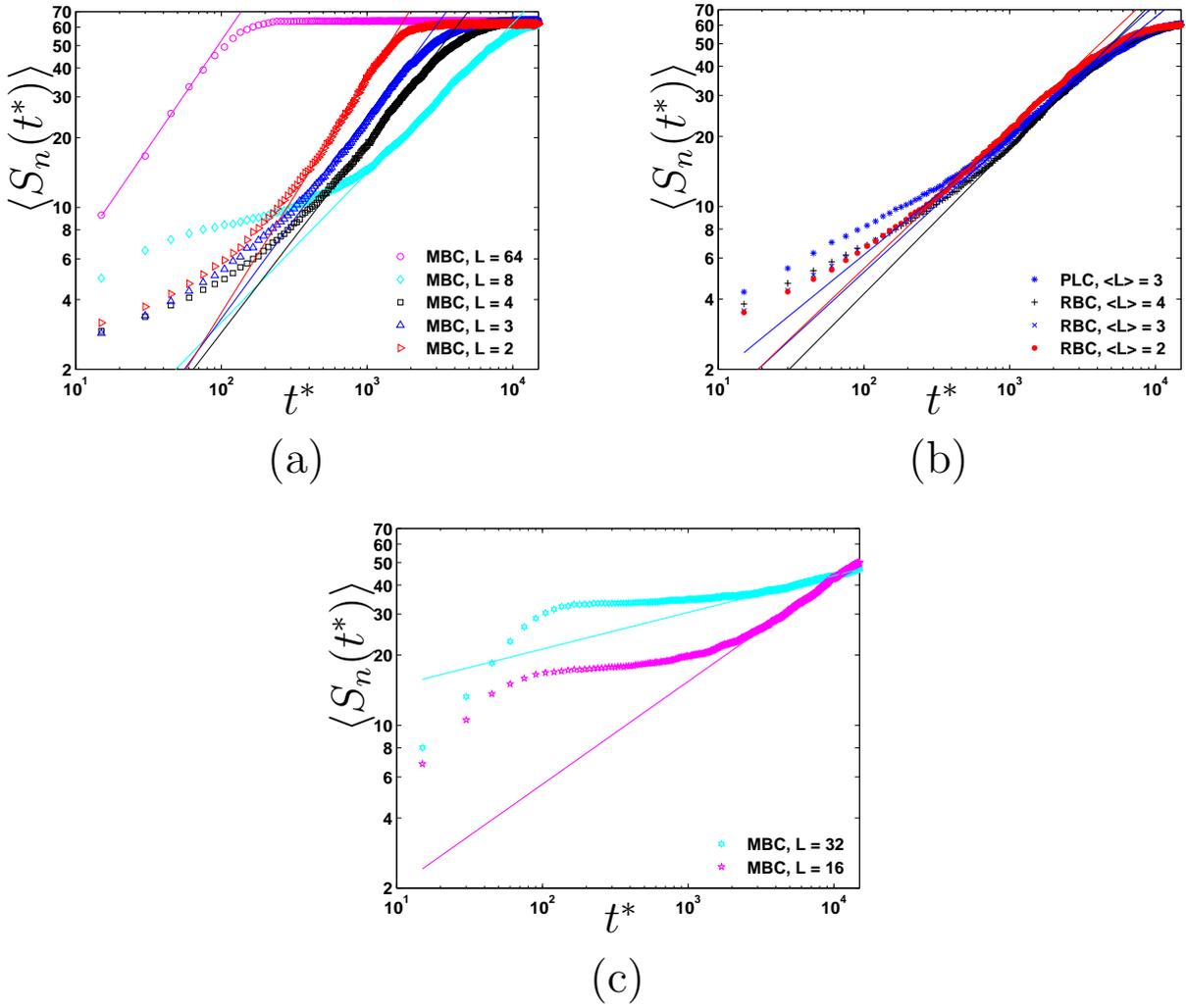}}
\vskip-40pt  
\caption{Variation of the number average cluster size with time for
  chains with $N=128$ in the presence of HI, for (a) MBC chains, (b)
  the PLC chain and RBC chains with $\avbr{L}\leq 4$, and (c) MBC
  chains with $L=32$ and 16.}
\label{fig:Snn120vstHI}
\vskip-40pt
\end{figure}


\begin{figure}[!tbp]
\centerline{\includegraphics[width=12cm,height=!]{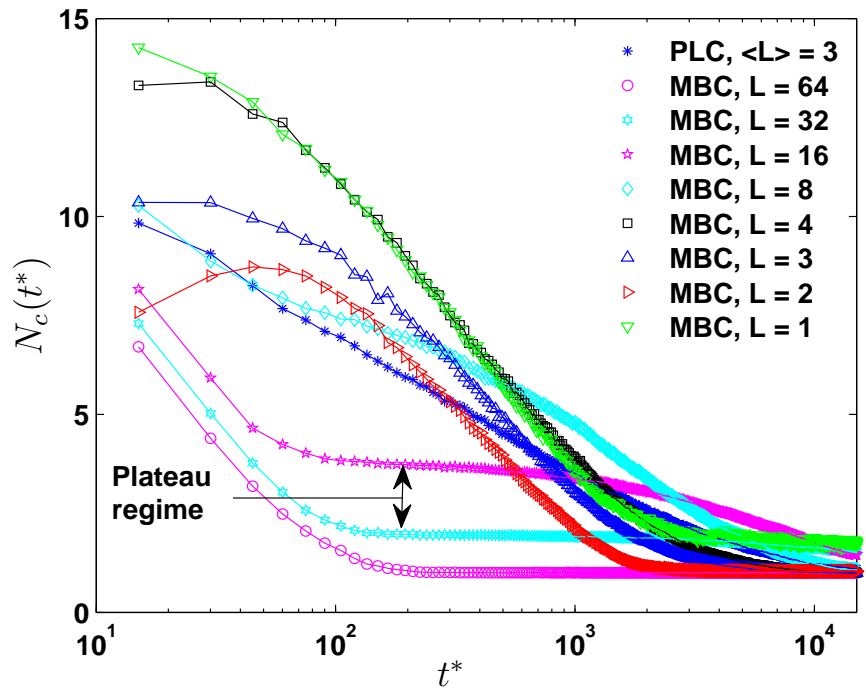}}
\vskip-40pt  
\caption{Variation of the number of clusters with time for the PLC and
  MBC chains, in the presence of HI.}
\label{fig:Ncnn120vstHI}
\vskip-20pt
\end{figure}


\begin{figure}[!tbp]
\centerline{\includegraphics[width=12cm,height=!]{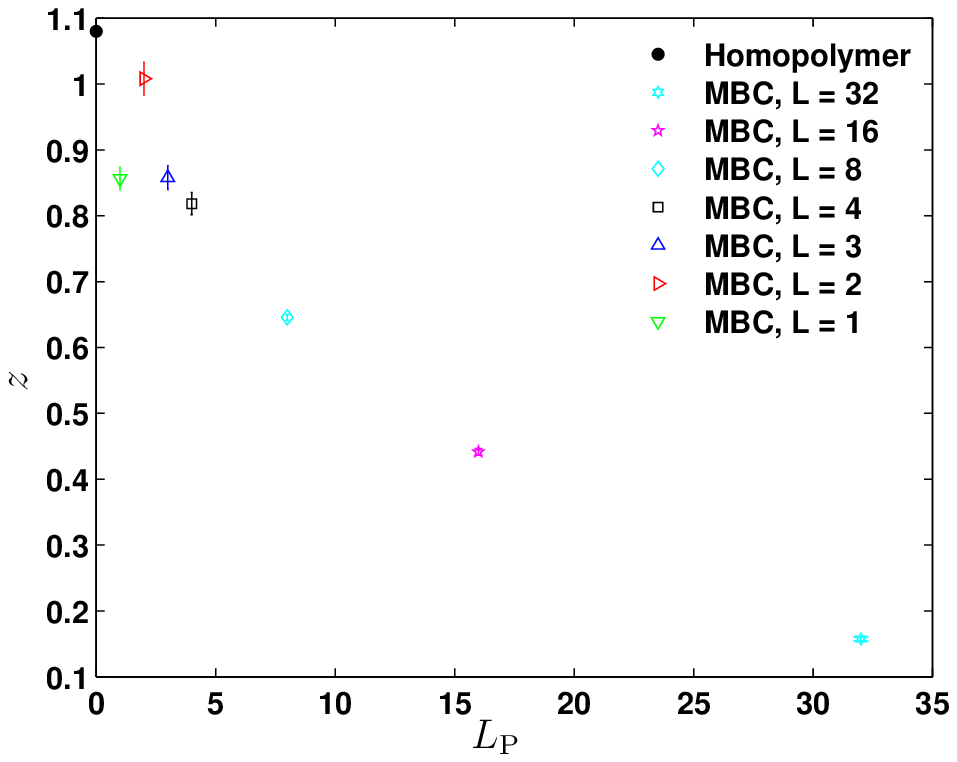}}
\vskip-40pt  
\caption{Coordinate pairs ($L_\text{P},z$) of the block size of P type
  monomers and the coarsening stage exponent, respectively, for MBC chains
  and a homopolymer chain with $N=128$, in the presence of HI.}
\label{fig:zvsLHI}
\vskip-20pt
\end{figure}

\subsection{\label{sec:Cluster} Cluster analysis}

The growth of the number average cluster size with time for our chosen
value of overlapping distance $D^* = 1.2 \times 2^{1/6} (\sigma /
l_H)$ is shown in Fig.~\ref{fig:Snn120vstHI}. The most striking aspect
about these data is their quantitative demonstration of the
multi-stage process of collapse kinetics, already indicated qualitatively 
by inspection of snapshots, in a much 
clearer fashion than by the data on the gyration radius. Typically, there
are three growth regimes: (i) Collapse of the single H blocks on a
fairly short time scale, giving rise to a moderate increase in the
mean cluster size $\avbr{S_n}$; (ii) Coalescence of
these clusters at a later time scale, resulting in a steep increase in
cluster size; and finally (iii) internal rearrangement of the
resulting ``sausage'' that does not change $\avbr{S_n}$ anymore.
However, for MBC chains with large block size (i.~e. $L=32$ and 16,
see Fig.~\ref{fig:Snn120vstHI} (c)) yet another (fourth) regime can be
observed, which follows directly the initial single-block collapse,
and which is characterised by a constant value of $\avbr{S_n}$. This
clearly corresponds to the time regime that is needed for the single
globules to find each other in a diffusion-like process. 

This diffusion regime for MBC chains with large P blocks is also
observed in the time evolution of the number of clusters $N_\text{c}$,
as shown in Fig.~\ref{fig:Ncnn120vstHI}: While the $L=64$ chain
(diblock copolymer) shows a simple decay of $N_\text{c}$ to
$N_\text{c} = 1$, i.~e. coalescence of smaller clusters to one large
globule, the chains with more blocks ($L = 32, 16, 8$) exhibit a clear
plateau at $N_\text{c} = N / (2L)$, i.~e. the number of H blocks. For
smaller block sizes, such a plateau is no longer clearly identifiable,
since the intervening P blocks are too short to keep the H clusters
away from each other. Interestingly, for $L = 2$, $N_\text{c}$
exhibits a clear maximum, which corresponds to breakup and coalescence
of clusters. Furthermore, the PLC chain, whose data is shown for
comparison, shows just a smooth decay to $N_\text{c} = 1$, without any
visible structure.

During the cluster coalescence stage, previous studies of homopolymer chains
indicate that one might expect a power law
\begin{equation}
\avbr{S_n(t)} \sim t^z . 
\end{equation}
We therefore fitted such a behaviour to our data; the results for the
exponent $z$ are listed in Table~\ref{tab:exponents}. Interestingly,
the $z$ values that we find for our copolymers are always smaller than
the value $z = 1.08 \pm 0.01$ obtained for a homopolymer
\cite{Phametal08b}. Given the fact that the obtained values differ
from each other substantially, it is far from clear that the obtained
numbers have much to do with a well-defined asymptotic power law.
Furthermore, the fact that the coalescence process involves two
competing length scales (the average globule size of the consolidated
clusters, and the size of the P ``loops'') makes the application of
self-similarity arguments, which are usually employed in order to
justify power-law behaviour, rather doubtful. On the other hand, the
fact that the $z$ value depends systematically on the block length (as
shown in Fig.~\ref{fig:zvsLHI} for the homopolymer and MBC chains up
to block length $L=32$) makes the interpretation of $z$ in terms of an
\emph{effective} exponent, which in reality describes a more
complicated dynamics, fairly likely.

Noticeably, data for MBC chains with $L=1$ and $L=2$ in Fig.~\ref{fig:zvsLHI} do not lie on the smooth curve that can be drawn by eye through all the remaining chains starting from the homopolymer chain to the MBC chain with $L=32$. As discussed earlier, for these two MBC chains, repulsive forces build up during the coalescence process due to the close proximity of the P blocks with each other. This leads to an equilibrium state with the highest energy (Fig.~\ref{fig:RgsqeqvsEVHI}), and the highest asphericity, as evident from the snapshot (Fig.~\ref{fig:MBCpath}), and the ratio of the largest to smallest eigenvalue (Fig.~\ref{fig:Eig1by3vstHI}). The non-occurrence of the final cluster consolidation stage into a compact globule as a consequence of the built up repulsive energy is probably the reason for the data for these two chains not following the general trend seen in the case of the other MBC chains. A similar pattern is seen below in the correlation between the collapse time and block size, and equilibrium mean square gyration radius.


\begin{figure}[tbp]
\centerline{\includegraphics[width=12cm,height=!]{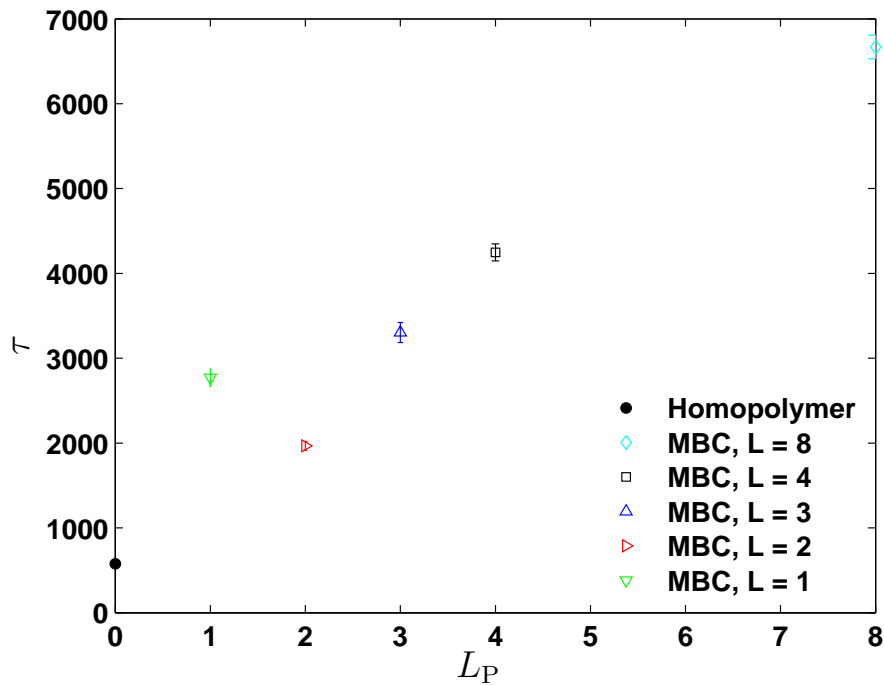}}
\vskip-40pt  
\caption{Coordinate pairs ($L_\text{P},\tau$) of the block
  size of P type monomers and total collapse time for MBC chains
  and a homopolymer chain with $N=128$, in the presence of HI.}
\label{fig:tauvsLHI}
\vskip-20pt
\end{figure}


\begin{figure}[tbp]
\centerline{\includegraphics[width=12cm,height=!]{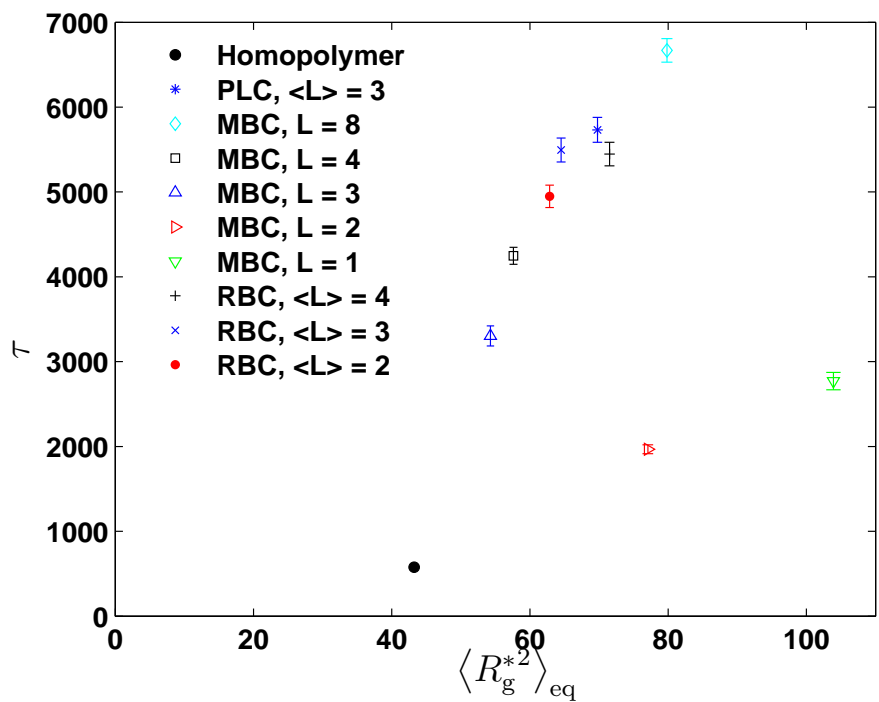}}
\vskip-40pt  
\caption{Total collapse time versus equilibrium mean square gyration
  radius for $N=128$ copolymer chains of different types, in the
  presence of HI.}
\label{fig:TauvsRgsqeqHI}
\vskip-20pt
\end{figure}

\subsection{\label{sec:CollTime} Collapse time}

The characteristic collapse times $(\tau)$ for all three types of
copolymers with $N=128$ are shown in Table~\ref{tab:exponents}. It is
observed that for all types of copolymers used in this work, the
collapse time is much larger than the collapse time for a homopolymer
chain, for which $\tau = 591 \pm 7$. As has been pointed out earlier,
MBC chains with $L=32$ and 16 had not yet reached their equilibrium
state during the course of our simulation, and hence the total
collapse time for these two chains are not available. The collapse
time also increases systematically with the block size, as
demonstrated in Fig.~\ref{fig:tauvsLHI} for the homopolymer and MBC
chains up to $L=8$. These data hence clearly indicate that the block
size of P monomers not only controls the duration of the diffusion
process, but also the collapse rate of the cluster coalescence stage
for copolymers with small P block size where diffusion plays little or
no role. The combined diffusion and cluster consolidation process is thus
identified as the decisive rate-limiting factor for the overall
collapse. The absence of these processes for a homopolymer leads to a 
much faster collapse, and also leads in the case of MBC chains with $L=1$ and $L=2$, to a departure 
from the observed trend for the other chains.

In order to understand the relationship between the kinetic
accessibility of the final state and the final equilibrium size, we
plot the collapse time versus the final size in
Fig.~\ref{fig:TauvsRgsqeqHI}. Except for 
MBC chains with $L=1$ and $L=2$ (for the reasons discussed earlier), the figure clearly
demonstrates that the total collapse time is directly related to the
final equilibrium size. This result reveals a very interesting feature
that is somewhat unexpected, i.~e., a chain with a small equilibrium
size tends to fold much more rapidly than a chain with a larger
equilibrium size. Intuitively, one might expect that it would take
longer for a chain to fold into a more compact equilibrium structure
rather than into a loosely packed structure, but the present results
suggest the opposite. In \Refcite{Salietal94} it was pointed out
that a pronounced energy minimum is a necessary condition to guarantee
that the native state is stable, and that such a minimum is
sufficient, for a compact globule with random structures, to rapidly
find the native state. Thus the deep energy minimum of the compact
structure seems to provide a guide (or a strong thermodynamic driving
force) for the chain to quickly fold into its final equilibrium
state. In the present work, we have made no attempts to determine the
free energy of the chains. However, we anticipate that knowledge of
the relationship between the free energy of the native state and the
size of the native conformation may help in understanding this
behaviour better.

\subsection{\label{sec:BlockSize} Block size distribution}


\begin{figure}[tbp]
\centerline{\includegraphics[width=12cm,height=!]{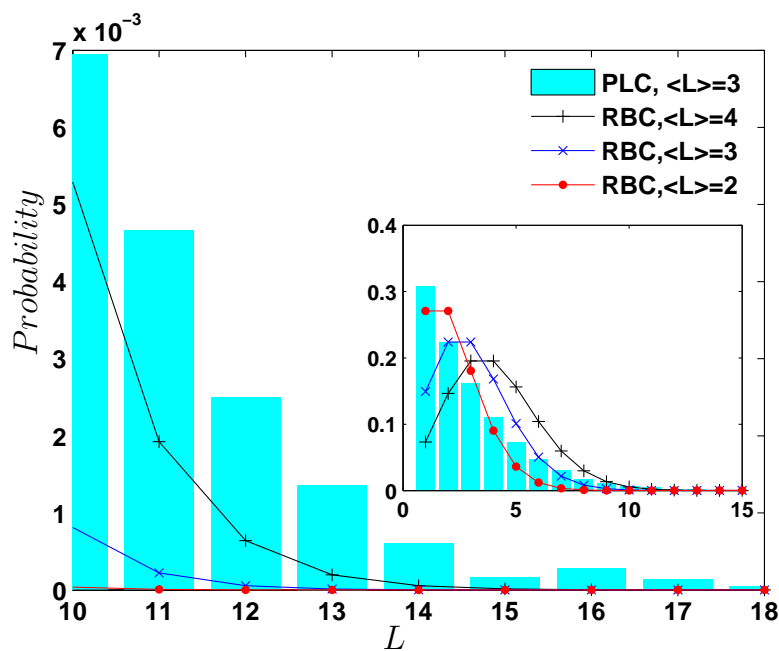}}
\vskip-40pt  
\caption{The tails of the normalised distribution of P blocks for a
  protein-like copolymer (PLC) and random-block copolymers (RBC) with
  three different average block lengths $\avbr{L}$, for $N=128$.
  Inset: The complete normalised distribution for PLC and RBC chains.}
\label{fig:Pblockdist}
\vskip-20pt
\end{figure}

As has been discussed above, the length of P blocks for MBC chains
has a very decisive influence on the kinetics (large blocks lead to a
substantial slowing down of the process) as well as on the equilibrium
size (large blocks increase the size). Since this influence is very
strong, one might expect that in disordered copolymers the presence of
such large blocks should also have a substantial influence, \emph{even
  if their probability of occurrence is fairly small}. We have
therefore analysed the probability distribution of P block lengths for
the PLC chains that we have generated, and compare it in
Fig.~\ref{fig:Pblockdist} to the Poisson distributions of RBC chains
with similar $\avbr{L}$. The inset shows the total distributions, but
for our purposes the tails of the distributions at large $L$ values
are of particular interest; hence they are shown in the main part. We
find the very interesting result that for the PLC chains the tail is
substantially enhanced compared to the Poissonians with $2 \le
\avbr{L} \le 4$. It is therefore quite natural to assume that it is
this enhanced tail that leads to the slowed-down dynamics of the PLC,
compared to the RBC with identical $\avbr{L}$. If this correlation is
really valid, then one should expect that the lattice PLC chains of
\Refcite{KhokhlovKhalatur99} should show a \emph{deflated} tail, in
comparison to the corresponding RBCs, as a result of different packing
behaviour. It should be quite interesting to test this.

\section{\label{sec:Conclusion} Summary and conclusions}

We have shown that Brownian Dynamics simulations incorporating
implicit hydrodynamic interactions can be used to study the dynamics
of copolymer collapse in a poor solvent. Our observations regarding
the speedup of collapse caused by hydrodynamic interactions, and the
existence of at least two stages of collapse, are similar to those
that have been reported previously in the literature. Due to the
inherent computational advantages of an implicit-solvent model,
compared to, say, straightforward Molecular Dynamics with explicit
solvent, we were able to study the phenomena with reasonable
statistical accuracy.

The kinetics of collapse can be described as a rapid initial formation
of clusters followed by cluster coalescence and sometimes a
rearrangement of the final cluster to form a compact state. It is also
found that the presence of P monomers pushes the value of quench depth
at which trapping phenomena occur to a higher value compared to the
value seen for homopolymer chains. A striking feature observed here is
that the total collapse time is completely governed by the cluster consolidation
stage and the rate of collapse of this stage depends on the block size
of P monomers along the chain. It is found that in our model random
block copolymers collapse more rapidly than ``protein--like''
copolymers with the same average block length, if the latter are
generated by the labelling procedure suggested in
\Refcite{KhokhlovKhalatur99}, and this behaviour can apparently be
traced back to an enhanced occurrence of long P blocks in the PLCs. 

We finally emphasise again that our model is intended as a ``toy
model'' designed to study the influence of certain aspects (monomer
sequence, hydrodynamic interactions) on the collapse kinetics, which
cannot be expected to be directly related to real proteins. To
illustrate this point, let us assume that each model monomer
corresponds to a single amino acid residue, with a typical size of,
say, $1 nm$, which would then be the value of $l_H$ in real
units. Since we are only interested in order--of--magnitude estimates
here, this could also be taken as the value for $\sigma$ or for
$Q_0$. A chain of $N = 64$ or $128$ would then correspond to a fairly
short polypeptide whose native state would typically need to be
described in terms of a much more elaborate geometry (e.~g. alpha
helices, beta sheets) than just a spherical or sausage--like globule.
The elementary time scale $\lambda_H$ is then found via
\begin{equation}
\lambda_H = \frac{\zeta}{4 H} =
6 \pi \eta_s a \frac{l_H^2}{4 k_B T} ,
\end{equation}
where $a$, the Stokes radius of the model monomer, is given by
\begin{equation}
a = h^\star \sqrt{\pi} l_H , 
\end{equation}
such that we find
\begin{equation}
\lambda_H = \frac{3}{2} \sqrt{\pi} h^\star \frac{\eta_s l_H^3}{k_B T} .
\end{equation}
For our value of $h^\star$, and assuming the viscosity of water and
room temperature, we thus find $\lambda_H \approx 0.3 ns$ as our
elementary time unit. A collapse time of $6000$ in our BD units would
thus correspond to roughly $2 \mu s$. This value should however not be
taken too seriously, since our model does not include bond--bending
and torsional potentials and other interactions that would be needed
to describe the structure and energetics on such a small length scale.
Furthermore, the value of $l_H$, which we have essentially just
guessed, enters with the third power. Nevertheless, it may be noted
that detailed simulations and experiments on real proteins
\cite{pande_nature} have also produced collapse times in the
microsecond range, whose values are however somewhat larger, for
chains that are even shorter. In general, we believe that the
simplicity of our model results in a lack of local kinetic barriers,
such that our estimate for the collapse time tends to be smaller than
the folding time of the ``corresponding'' real protein. In this
context, it should also be noted that the interpretation of one model
monomer in terms of just one residue is certainly not imperative. One
could as well consider a model where one model monomer corresponds to
several residues (under the assumption that proteins with such long
blocks exist). In this case, one would have a larger value for $l_H$,
and therefore a very much larger folding time in real units.

\section*{Acknowledgements}

This work was supported by the Australian Research Council under the
Discovery Projects program. Computational resources were provided by
the Australian Partnership for Advanced Computation (APAC), Victorian
Partnership for Advanced Computation (VPAC) and the Monash Sun Grid
Cluster. Tri Pham acknowledges hospitality at the Mainz Max Planck
Institute for Polymer Research within the framework of the
International Max Planck Research School.


\providecommand{\refin}[1]{\\ \textbf{Referenced in:} #1}

\end{document}